\documentclass[prd, preprint, nofootinbib, showpacs]{revtex4}

\usepackage{epsf,epsfig,subfigure,graphicx,amsmath,amssymb}
\usepackage{amsfonts}
\usepackage{latexsym}
\usepackage{color}

\newcommand{\dis}[1]{\begin{equation}\begin{split}#1\end{split}\end{equation}}
\newcommand{\be}{\begin{equation}}
\newcommand{\ee}{\end{equation}}
\def\bea{\begin{eqnarray}}
\def\eea{\end{eqnarray}}

\newcommand{\eq}[1]{Eq.~(\ref{#1})}

\newcommand{\bfrac}[2]{{\left(\frac{#1}{#2} \right)  }}

\newcommand{\Mp}{M_P}

\newcommand\gev{\,{\rm GeV}}

\newcommand{\calP} {{\cal P}}
\newcommand{\tilder} {{\tilde r}}

\newcommand{\fnl}{f_{\rm NL}}
\newcommand{\gnl}{g_{\rm NL}}
\newcommand{\taunl}{\tau_{\rm NL}}

\begin{document}

\title{
Can Standard Model Higgs Seed the Formation of Structures in Our Universe?
}

\author{Ki-Young Choi$^{1}$\footnote{kiyoung.choi@apctp.org} and Qing-Guo Huang$^{2}$\footnote{huangqg@itp.ac.cn}}
\affiliation{${}^{1}$ Asia Pacific Center for Theoretical Physics, Pohang, Gyeongbuk 790-784, Republic of Korea and \\
Department of Physics, POSTECH, Pohang, Gyeongbuk 790-784, Republic of Korea}

\affiliation{${}^{2}$ State Key Laboratory of Theoretical Physics, Institute of Theoretical Physics, Chinese Academy of Science, Beijing 100190, People's Republic of China}

%

\begin{abstract}
We study the Standard Model Higgs field as a source for the primordial curvature perturbation, particularly  in 
the curvaton and modulated reheating scenario. We conclude that the Higgs cannot play as a curvaton due to the small energy density when it decays,  however the modulated reheating by Higgs can be a 
viable scenario.  In the latter case, the non-Gaussianity is inevitably generated and 
strongly constrains the type of potential of inflaton field and Higgs-dependent interaction term.
For the quadratic potential of the inflaton field with  decay rate  which  non-linearly depends on the Higgs vacuum expectation value, the  contribution  of Higgs field to the primordial curvature perturbation must be less than 8 \%.

\end{abstract}

\pacs{98.80.-k,98.80.Cq,98.80.Es}

\preprint{APCTP-Pre2012 - 015} 

\vspace*{3cm}
\maketitle


\section{Introduction}
Cosmic inflation solves various problems
 in the standard Big Bang cosmology~\cite{Inflation}
 and simultaneously
 provides the seed of large scale structure in our Universe from
 the quantum fluctuation of a light scalar field~\cite{InflationFluctuation}.

Recently ATLAS and CMS collaborations at CERN  reported a discovery of Higgs-like particle
with the mass of 125 $\gev$~\cite{:2012gk,:2012gu}.  As a unique scalar field in the Standard Model (SM), Higgs
can be considered as a source field for the inflation.
However it is known that  its large self-interaction predicts very large  density perturbation, which is easily ruled out by observation~\cite{quarticinflation,Isidori:2007vm}. 

It was recently demonstrated~\cite{CervantesCota:1995tz,Bezrukov:2007ep} that the SM itself can 	give rise to inflation, provided non-minimal coupling of the Higgs field with gravity. However this model is plagued by several issues such as unitarity problem~\cite{unitarity}, and the stability of the potential up to the near Planck scale~\cite{Isidori:2007vm}. 
The sensitivity of the Higgs inflation scenario to the details on the UV completion of the theory was studied recently in~\cite{Bezrukov:2010jz}.

Even though the canonical Higgs  field in the Standard Model is not a good candidate for inflation, it may contribute to the primordial curvature perturbation to seed the structure formation and anisotropies in cosmic microwave background radiation. The Higgs perturbation is generated during inflation on super horizon scales at horizon exit if its mass is smaller than the Hubble scale. During inflation this remains the isocurvature perturbation, however it can be converted to the adiabatic one after or at the end of inflation.
In this case Higgs can contribute to the primordial curvature perturbation.

In the Ref.~\cite{DeSimone:2012qr}, the authors studied the possibility that the SM Higgs might be responsible for the inhomogeneties we observe in our universe in the alternative scenarios such as modulated decay~\cite{Dvali:2003em} and inhomogeneous end of inflation~\cite{Lyth:2005qk,Lyth:2006nx}.
In particular they discussed about the implications for the detection of primordial tensor perturbations in the future observation of $B$-mode of CMB polarization.

In this paper we  study the idea of the Higgs field as a source for the primordial curvature perturbation,
however we focus on the observables from the non-linear terms in the curvature perturbations, the non-Gaussianity.   Wilkinson Microwave Anisotropy Probe (WMAP) seven-year data constrains the local type nonlinear parameter,
 $-10<\fnl<74$ at the 95\% confidence level~\cite{Komatsu:2010fb}. 

 We find that Higgs as a curvaton is difficult since the energy density of the Higgs condensate is too small at the time of energy transfer of Higgs to the radiation. In the modulated reheating scenario, Higgs can  contribute to the primordial curvature perturbation as shown in~\cite{DeSimone:2012qr}. However  we find that the Higgs-dependent decay can generate large non-Gaussianity. For the chaotic inflation with quadratic potential, too large non-Gaussianity is generated to be compatible with the current observation due to the non-linearly Higgs-dependent decay. In this case, we find that the Higgs contribution  to the primordial curvature perturbation should be less than  $8 \%$. 

The paper is organized as follows. In Section~\ref{higgs} we summarize the properties of Higgs field during inflation.  In Section~\ref{curvaton} we study the possibility of Higgs as curvaton.
In Section~\ref{modulated} we consider the modulated reheating by Higgs. In Section~\ref{discussion}
we summarize our results.

\section{Higgs field during inflation}
\label{higgs}

We consider the SM Higgs potential in the form,
\be
V(h)=-{1\over 2} m^2 h^2+{\lambda \over 4}h^4, \label{HiggsPotential}
\ee
where the mass of Higgs in the minimum is $m_h= \sqrt{2} m \simeq 125$ GeV and the self coupling $\lambda\equiv \lambda(\mu)$ has a logarithmic dependence on the energy scale.
 When we use the central values for the top quark mass and strong coupling constant, the Higgs potential develops an instability around $10^{11}\gev$, with a lifetime much longer than the age of the Universe. However taking into account theoretical and experimental errors, stability up to the Planck scale cannot be excluded~\cite{EliasMiro:2011aa,Bezrukov:2012sa,Degrassi:2012ry}. 
Therefore in this paper it is legitimate to assume that the Higgs potential is stable up to the energy scale of our interest below Planck scale.
  
Since the canonical Higgs field is not suitable for inflation, we assume that there is another scalar field which drives inflation  at very high energy scale. The scale of the inflation is constrained by the non-observation of the tensor spectrum.  The present bound on the tensor-to-scalar ratio gives upper bound on the Hubble parameter during inflation~\cite{Komatsu:2010fb},
\dis{
H \lesssim 3\times 10^{14} \gev \label{Hstar}.
}
At this energy scale of inflation which is much larger than the electroweak scale,  it is reasonable to simplify Higgs potential \eq{HiggsPotential} to be quartic as
\dis{
V(h) \simeq \frac{\lambda}{4} h^4,
}
where  the size of the Higgs self coupling is $\lambda(\mu)\simeq \mathcal{O} (10^{-2})$~\cite{EliasMiro:2011aa,Degrassi:2012ry}. For simplicity we take $\lambda \simeq 0.01$ and constant at the scale of our study.

For the generation of the quantum fluctuations of the Higgs field, we require that the effective mass of the Higgs,\dis{
m^2_{\rm eff}= V_{hh} = 3\lambda h^2, 
}
is smaller than the Hubble parameter during inflation.  This condition sets the upper bound on the Higgs VEV
\dis{
h \ll \frac{H}{\sqrt{3\lambda}} \lesssim  2\times10^{15}\gev, \label{cond_h}
}
where we used \eq{Hstar} and $\lambda = 0.01$.
Because Higgs is quite light~\footnote{The Higgs is plagued by the quadratic divergence of its mass, but in our discussion we assume that the hierarchy problem is solved in some way and we don't address this problem.}, it almost does not move from its initial position and stays there during inflation. At this stage, one can easily check that the Higgs energy density is subdominant compared to the inflaton energy density by the order of less than $H^2/\Mp^2$ as long as \eq{cond_h} is satisfied.   

On the other hand, the Higgs  fluctuations is generated during inflation with the amplitude
\dis{
\delta h = \frac{H}{2\pi}.
}
Combining with \eq{cond_h}, we lead to the range of the Higgs perturbation to Higgs VEV 
\dis{
0.03 \simeq \frac{\sqrt{3\lambda}}{2\pi} \lesssim \frac{\delta h }{h}  \lesssim 1,
\label{dhh}
}
which is independent on the precise value of Hubble parameter.
Here the upper bound comes by considering that the  VEV of Higgs should be larger than the amplitude of its quantum fluctuations. 

\section{Higgs field as curvaton}
\label{curvaton}
A light scalar field which is subdominant during inflation can decay very late after inflation, when its vacuum energy is transfer to the radiation. At this time the isocurvature perturbation of curvaton field can be converted to the adiabatic one and contribute to the primordial curvature perturbation. This is known as curvaton scenario~\cite{Mollerach:1989hu,Linde:1996gt,CurvatonLW,CurvatonMT,CurvatonES}. There have been many studies on this with different types of potentials of curvaton~\cite{CurvatonSelf}  including the inflaton contribution~\cite{MixedIC} to study the power spectrum and non-Gaussianities~\cite{CurvatonNG}. 
Since Higgs is subdominant during inflation and decay late it can be a natural candidate for curvaton.
In this section we study the possibility of the SM Higgs field  as  curvaton.

Inflation is assumed to be driven by another scalar field, inflaton. During inflation the energy density of Higgs is  subdominant and it acquires  perturbation  as explained in section~\ref{higgs}. 
After inflation, the inflaton field oscillates around its local minimum and its potential energy is converted into radiations
to initiate the radiation-dominated Universe. 
During this time, Higgs field starts oscillation when the effective mass becomes bigger than the Hubble expansion, namely when
\dis{
H^2_{\rm osc} =m^2_{\rm eff}= 3\lambda h^2.
}
Since the potential of Higgs is quartic, 
the energy density of an oscillating Higgs decreases as $a^{-4}$ with scale factor $a$ which is similar to radiations. 
Therefore during the epoch of the oscillation of Higgs, the ratio of Higgs energy density to that of the radiation-dominated  background continues to be constant.
Actually, after some oscillations of Higgs, the energy of Higgs condensate is transferred to the radiation by perturbative and non-perturbative processes.
Since the coupling of Higgs to other fields are strong the energy transfer is completed soon after the start of oscillation~\cite{Bezrukov:2008ut,GarciaBellido:2008ab}. 
Therefore the Higgs density parameter $\Omega_h$ at the time of Higgs decay is approximatetely the same
as the onset of oscillation of Higgs, which is given by
\dis{
\Omega_h=\frac{\rho_h}{\rho_r} = \left( \frac{\rho_h}{\rho_r}\right)_{\rm osc} = \frac{\lambda/4\, h_{\rm osc}^4}{3\Mp^2 H_{\rm osc}^2} =\frac{h^2}{36\Mp^2} \ll \frac{H^2}{108 \lambda \Mp^2} \lesssim  1.5\times 10^{-8}.
\label{omegah}
}
Here we used  $h_{\rm osc}=h $, the higgs value during inflation and  the inequality comes using \eq{Hstar} and \eq{cond_h}. 
The density perturbation generated  after the decay of Higgs depends on the energy density of the Higgs when it decay, and it is given by  
\be
{\delta \rho\over \rho}\sim {\rho_h\over \rho}{\delta \rho_h\over \rho_h}\sim {\rho_h\over \rho} {\delta h\over h}\sim \Omega_h {\delta h\over h}.
\ee
The Higgs perturbation $\delta h/h$ is already bounded to be less than ${\mathcal O}(1)$ in~\eq{dhh}.
Therefore we can easily see that, due to the small energy density of Higgs in~\eq{omegah}, it cannot produce enough density perturbation which is necessary for  observation today. Thus we conclude that Higgs cannot play as curvaton.
In the Ref.~\cite{Kunimitsu:2012xx}, the authors also argued that Higgs is not viable as curvaton  in the context of generalized G inflation model.

\section{Modulated reheating by Higgs }
\label{modulated}
After inflation the energy density in the inflaton field must be transferred into radiation. In the simplest case of 
single field inflation model, reheating process does not affect the primordial curvature perturbation on scales which are observable today, because these scales were much larger than the horizon at the time of reheating. 
However when there is  a subdominant light scalar field which modulates the efficiency of the reheating 
the situation changes~\cite{Dvali:2003em,Kofman:2003nx,Zaldarriaga:2003my}.  The quasi-scale invariant perturbations in this light field, which during inflation are an isocurvature perturbation, can be converted into the primordial curvature perturbation during this process~\cite{Alabidi:2010ba}. 
In this section, we examine the possibility of Higgs as a modulating particle in the reheating after inflation.

We assume that the inflaton field, $\phi$, drives inflation and its subsequent decay during oscillation transfers its vacuum energy  to the background  radiation.
We consider the Lagrangian
\dis{
{\mathcal L} = \frac12 \partial_\mu \phi \partial^\mu \phi -V(\phi) + { \mathcal L_{\rm int}},
}
where $V(\phi)$ is the potential which is responsible for the inflation and ${ \mathcal L_{\rm int}}$ is the interaction  responsible for the inflaton decay, which depends on the Higgs VEV, $h$.
For simplicity we will take a polynomial potential for inflaton, $V(\phi) \propto \phi^{2\alpha}$. For the Higgs dependent  interactions we consider~\cite{Ichikawa:2008ne}
\dis{
{\mathcal L}_{\rm int}  \supset -\sum_a y_a(h) \phi \bar{\psi}_a \psi_a - \sum_a M(h) \phi \chi_a^2 -\sum_a g_a(h)\phi^2\chi_a^2, \label{Lint}
}
where $\chi_a$ and $\psi_a$ are scalar and fermion fields which contribute to the radiation in the early Universe.
The coupling constants $y_a(h), M_a(h)$, and $g_a(h)$ are functions of the Higgs field $h$.

The total decay rate of the inflaton can be composed of Higgs independent and  dependent part,
\begin{equation}
 \Gamma (h)=  \Gamma^{I} + \Gamma^{D}(h).
\label{Decay:total}
\end{equation}
Using \eq{Lint}, Higgs dependent decay rate  $\Gamma^{D}(h)$ has the form of~\cite{Ichikawa:2008ne}
\dis{
\Gamma^{D}(h)=\sum_a A_n \frac{y_a^2(h)}{8\pi} m_\phi^{\rm eff} + \sum_a B_n \frac{M_a(h)^2}{8\pi m_\phi^{\rm eff}} + \sum_a C_n \frac{h_a^2(h)}{8\pi (m_\phi^{\rm eff})^3} \rho_\phi,
}
with $m_\phi^{\rm eff}$ the effective mass of the inflaton field defined by $(m_\phi^{\rm eff} )^2=\partial^2 V/ \partial^2\phi $ and $A_n,B_n$, and $C_n$ are numerical coefficients of the order of ${\mathcal O}(1-1000)$~\cite{Ichikawa:2008ne}. 

After the reheating process, 
the primordial curvature perturbation $\zeta$ has two contributions from the inflaton and the Higgs field.
Due to the slow-rolling the inflaton field does not contribute to the curvature perturbation at non-linear orders.
In contrast the modulating field can easily generate the non-linear contribution to the $\zeta$.
In this  mixed modulating scenario, the $\zeta$ can be written as~\cite{Zaldarriaga:2003my, Ichikawa:2008ne} 
\begin{eqnarray}
\zeta= \frac{1}{\Mp^2}\frac{V}{V_\phi}\delta\phi_* +Q_h\delta h_* + \frac12Q_{hh}\delta h_*^2
 + \frac{1}{6}Q_{hhh}\delta h_*^3 +\cdots ,
\label{zeta}
\end{eqnarray}
where $Q$ is a function of $\Gamma(h)/H_c$ calculated  at a time $t_c$ which  is after several oscillations of the inflaton but well before the time of decay of inflaton. A quantity with subscript $*$ is evaluated  when the corresponding scale crosses the Hubble horizon during inflation. For $\Gamma/H_c \ll 1$, $Q$ can be well approximated by~\cite{Zaldarriaga:2003my, Ichikawa:2008ne} 
\dis{
Q = a_0 \log \left(  \Gamma / H_c \right),
}
where $a_0$ has different value and sign for different inflaton potential and  interaction for the dominant decay mode. 
The value of $a_0$ for some of the examples are given in the Table I of Ref.~\cite{Ichikawa:2008ne}.  
With this form we find the derivatives
\dis{
Q_h &= a_0 \frac{\Gamma_h}{\Gamma},\\
Q_{hh} &= a_0 \left(  \frac{\Gamma_{hh}}{\Gamma} - \frac{\Gamma^2_h}{\Gamma^2}\right), \\
Q_{hhh} &= a_0  \left(  \frac{\Gamma_{hhh}}{\Gamma} - 3\frac{\Gamma_h \Gamma_{hh}}{\Gamma^2} + 2\frac{\Gamma^3_h}{\Gamma^3}  \right). \label{Qderivatives}
}
Once $\zeta$ in~\eq{zeta} is given,
we can easily calculate  the power spectrum of curvature perturbation
\dis{
 {\cal P}_{\zeta} = \calP_{\zeta_\phi} +\calP_{\zeta_h} =\frac{1}{2\Mp^2\epsilon_*} \bfrac{H_*}{2\pi}^2  (1+\tilder).\label{Pzeta}
}
The power spectrum has  contributions from each field and we define the ratio by $\tilder$ as
\begin{equation}
\tilder \equiv \frac{{\cal P}_{\zeta_h}}{{\cal P}_{\zeta_\phi}}= 2\Mp^2 \epsilon_*Q_h^2, 
\end{equation}
where $\epsilon$ is the slow-roll parameter
\be
\epsilon\equiv \frac{\Mp^2}{2}\bfrac{V_\phi}{V}^2. 
\ee
The spectral index of power spectrum of curvature perturbation is also obtained
\be
n_s\equiv 1+{d\ln {\cal P}_\zeta\over d\ln k}=(1-\beta) n_{\zeta_\phi}+\beta n_{\zeta_h}, 
\ee
where $\beta\equiv \tilder /(1+\tilder)\in [0,1]$.
Here we used the notations
\bea
n_{\zeta_\phi}&=& 1-6\epsilon+2\eta_{\phi\phi}, \\
n_{\zeta_h}&=& 1-2\epsilon+2\eta_{hh}, 
\eea
and 
\be
\eta_{\phi\phi} \equiv \Mp^2 \frac{V_{\phi\phi}}{V}, \quad 
\eta_{hh}\equiv {V_{hh}\over 3H^2}. 
\ee

The tensor perturbation  depends only on the total energy density during inflation, and thus the amplitude of its power spectrum takes the form 
\be
{\cal P}_T={H_*^2/\Mp^2\over \pi^2/2}, 
\ee
whose tilt is given by 
\be
n_T\equiv {d\ln{\cal P}_T\over d\ln k}=-2\epsilon_*. 
\ee
Usually we introduce a new parameter, so-called tensor-to-scaler ratio $r$, to measure the size of gravitational wave perturbation, 
\be
r\equiv {{\cal P}_T\over {\cal P}_\zeta}=(1-\beta)16\epsilon_*. 
\ee 
And thus we obtain the consistency relation  scenario between $n_T$ and $r$ for inflaton-modulated reheating,
\be
n_T=-{1\over 1-\beta}{r\over 8}. 
\ee
In the limit of $\beta\rightarrow 0$, the consistency relation becomes the prediction of the usual single-field slow-roll inflation model.

Without loss of generality, we can assume that $\Gamma^D(h)$ is proportional to polynomial of the Higgs field, $h^n$, with $n=1,2,3,\cdots$. For each case of $n$, $Q_h$ in \eq{Qderivatives} takes the form
\dis{
Q_h = a_0 B_h \frac{n}{h},\label{Qh_poly}
}
where $B_h \equiv \Gamma^D / \Gamma$, the fraction of the Higgs-dependent decay rate to the total decay rate of inflaton field. 
Using the observed power spectrum, 
$\calP_\zeta=2.46\times 10^{-9}$~\cite{Komatsu:2010fb}, and \eq{Pzeta}, then $B_h$ has to satisfy the relation 
\be
B_h= 3.1\times 10^{-4} {\sqrt{\beta}\over n|a_0|} \bfrac{h_*}{H_*} \ll 1.7\times 10^{-3} {\sqrt{\beta}\over n|a_0|},
\label{bbh}
\ee
where the last inequality comes from \eq{cond_h}.
Considering that $\sqrt{\beta}\sim {\cal O}(10^{-1}\sim 1)$ and $a_0 \sim 0.1$, we conclude that 
\be
B_h\lesssim {\cal O}(10^{-3}\sim 10^{-2}). 
\ee
It indicates that inflaton decay is dominated by the Higgs independent part.

During the modulated reheating process by Higgs, the non-linear term of  $\zeta$ in \eq{zeta} can generate large non-Gaussianity, which might be inconsistent with observational bound.
From \eq{zeta},  we can  calculate the non-linearity parameters $\fnl$, $\taunl$, and $\gnl$. Following~\cite{Ichikawa:2008ne},  in the leading order of the slow-roll parameters, they are
\dis{
\fnl &= \frac56 \frac{\beta^2}{a_0} \left( -1 + \frac{\Gamma\Gamma_{hh}}{\Gamma_h^2} \right),\\
\taunl&= \frac{\beta^3}{a_0^2}\left(1- \frac{\Gamma\Gamma_{hh}}{\Gamma_h^2} \right)^2,\\
\gnl&= \frac{25\beta^3}{54a_0^2}\left(2- 3\frac{\Gamma\Gamma_{hh}}{\Gamma_h^2} + \frac{\Gamma^2\Gamma_{hhh}}{\Gamma_h^3} \right).\\
}
Taking into account of the polynomial form of the Higg-depenent decay rate, $\Gamma^D(h)\propto h^n$, $f_{\rm NL}$ can be written by 
\dis{
\fnl 
\simeq -{5\over 6}{\beta^2\over a_0}+\frac56 \frac{\beta^2}{a_0B_h}\frac{n-1}{n}. 
\label{fnlh}
}
When the linear term dominates the Higgs-dependent term  ($n=1$), only the first term survives and  $\fnl\simeq -{5\over 6}{\beta^2\over a_0}$, which is smaller than ${\mathcal O}(1\sim 10)$. This is consistent with observation.  However when the non-linear term dominates ($n>1$), $f_{\rm NL}$ is mostly determined by the second term in \eq{fnlh} and can be large due to the small $B_h$. 
For this case with large $\fnl$, the other non-Gaussianity parameters  become 
\bea
\taunl &\simeq& \frac{36}{25\beta} \fnl^2,\\
\gnl&\simeq& {2(n-2)\over 3(n-1)\beta}\fnl^2-{5\over 3}{\beta\over a_0}\fnl.
\eea
If $\beta\ll 1$, $\tau_{\rm NL}$ are significantly enhanced compared to $\fnl^2$.

 For $n=1$, we have $a_0=-1/6$  if the inflaton potential is quadratic, $V(\phi)\propto \phi^2$~\cite{Ichikawa:2008ne}. The non-lineariy parameter is $\fnl=5\beta^2\leq 5$ and then the current bound on $\fnl$ does not give any constraint on $\beta$ \footnote{If a positive $\fnl>5$ is detected by forthcoming observations, such as PLANCK satellite~\cite{Planck}, this case can be probed. } ; if $V(\phi)\propto \phi^6$ we have $a_0=1/6,\ 1/30,\ 1/18$ corresponding to the dominant interaction  for inflaton decay $-y\phi \bar\psi \psi$, $-M\phi \chi^2$, $-\lambda_\phi\phi^2\chi^2$ respectively, and the bound on $\fnl$ implies 
\be
\beta<{\rm min} (\sqrt{12 a_0},1). 
\ee
We see that the non-Gaussianity does not constrain the model and the Higgs can significantly contribute to the curvature perturbation in the case of $n=1$. 

However when the non-linearly dependent term dominates the result changes.
For $n>1$, 
with the quadratic type of inflaton potential, $V(\phi)\propto \phi^2$,  the present bound on the nonlinear parameter $\fnl$ gives strong constraint. We obtain the upper bound   $\beta^2<{12n\over n-1}|a_0|B_h$ for negative $a_0$. 
Combining  with \eq{bbh}, this gives 
\be
\beta\lesssim {0.075\over (n-1)^{2/3}}. \label{betabound}
\ee
Since $\beta\simeq \tilder$ for small $\beta$, the constraint on $\beta$ translates into the bound on the contribution of the Higgs field to the  power spectrum. From \eq{betabound}, we conclude that the Higgs can contribute only  $8\%$ to the power spectrum of the primordial curvature perturbation and this is very subdominant.


\section{discussion}
\label{discussion}

We have examined the viability of  the Standard Model Higgs as a dominant source of the primordial curvature perturbation. We find that Higgs as a curvaton is difficult since the energy density of the Higgs condensate is too small at the time of energy transfer of Higgs to the radiation. In the modulated reheating scenario, Higgs can dominantly contribute to the perturbation when the Higgs-dependent decay rate is linearly proportional to the Higgs VEV. In this case the primordial non-Gaussianity generated by Higgs must be small. On the other hand, if the Higgs-dependent decay rate  depends non-linearly on the Higgs VEV,  too large non-Gaussianity is predicted. The current bound on the non-linearity parameter shows that  the contribution of the Higgs to the total curvature perturbation must be smaller than  $8\%$  for the quadratic potential of  inflaton field.

\section*{Acknowledgments}
K.-Y.C was supported by Basic Science Research Program through the National Research Foundation of Korea (NRF) funded by the Ministry of Education, Science and Technology (No. 2011-0011083).
K.-Y.C acknowledges the Max Planck Society (MPG), the Korea Ministry of
Education, Science and Technology (MEST), Gyeongsangbuk-Do and Pohang
City for the support of the Independent Junior Research Group at the Asia Pacific
Center for Theoretical Physics (APCTP).
K.-Y.C would like to thank the ITP of CAS for warm hospitality during his stay
 where this work was initiated. 
QGH is supported by the project of Knowledge Innovation Program of Chinese Academy of Science and a grant from NSFC (grant NO. 10975167).




\begin{thebibliography}{99}

\bibitem{Inflation}
  A.~A.~Starobinsky,
  JETP Lett.\  {\bf 30} 682 (1979)
  [Pisma Zh.\ Eksp.\ Teor.\ Fiz.\  {\bf 30} 719 (1979)];\\
  K.~Sato,
  Mon.\ Not.\ Roy.\ Astron.\ Soc.\  {\bf 195}, 467 (1981);\\
  A.~H.~Guth,
  Phys.\ Rev.\  D {\bf 23}, 347 (1981); \\
  A.~D.~Linde,
  Phys.\ Lett.\  B {\bf 108} 389 (1982);\\
  A.~Albrecht and P.~J.~Steinhardt,
  Phys.\ Rev.\ Lett.\  {\bf 48} 1220 (1982).

\bibitem{InflationFluctuation}
  S.~W.~Hawking,
  Phys.\ Lett.\  B {\bf 115}, 295 (1982); \\
  A.~A.~Starobinsky,
  Phys.\ Lett.\  B {\bf 117}, 175 (1982); \\
  A.~H.~Guth and S.~Y.~Pi,
  Phys.\ Rev.\ Lett.\  {\bf 49}, 1110 (1982).

\bibitem{:2012gk}
  G.~Aad {\it et al.}  [ATLAS Collaboration],
  [arXiv:1207.7214 [hep-ex]].
\bibitem{:2012gu}
  S.~Chatrchyan {\it et al.}  [CMS Collaboration],
  Phys.\ Lett.\ B
  [arXiv:1207.7235 [hep-ex]].
  
  
\bibitem{quarticinflation}
  A.~D.~Linde,
  Chur, Switzerland: Harwood (1990) 362 p. (Contemporary concepts in physics, 5)
  [hep-th/0503203],
  D.~H.~Lyth and A.~Riotto,
  Phys.\ Rept.\  {\bf 314} (1999) 1
  [hep-ph/9807278],
  A.~D.~Linde,
  Lect.\ Notes Phys.\  {\bf 738} (2008) 1
  [arXiv:0705.0164 [hep-th]].
  
\bibitem{Isidori:2007vm}
  G.~Isidori, V.~S.~Rychkov, A.~Strumia and N.~Tetradis,
  Phys.\ Rev.\ D {\bf 77} (2008) 025034
  [arXiv:0712.0242 [hep-ph]].

  


\bibitem{CervantesCota:1995tz}
  J.~L.~Cervantes-Cota and H.~Dehnen,
  Nucl.\ Phys.\ B {\bf 442} (1995) 391
  [astro-ph/9505069].
  
\bibitem{Bezrukov:2007ep}
  F.~L.~Bezrukov and M.~Shaposhnikov,
  Phys.\ Lett.\ B {\bf 659} (2008) 703
  [arXiv:0710.3755 [hep-th]].

\bibitem{unitarity}
  C.~P.~Burgess, H.~M.~Lee and M.~Trott,
  JHEP {\bf 0909} (2009) 103
  [arXiv:0902.4465 [hep-ph]],
  J.~L.~F.~Barbon and J.~R.~Espinosa,
  Phys.\ Rev.\ D {\bf 79} (2009) 081302
  [arXiv:0903.0355 [hep-ph]],
  C.~P.~Burgess, H.~M.~Lee and M.~Trott,
  JHEP {\bf 1007} (2010) 007
  [arXiv:1002.2730 [hep-ph]].

  
  
\bibitem{Bezrukov:2010jz}
  F.~Bezrukov, A.~Magnin, M.~Shaposhnikov and S.~Sibiryakov,
  JHEP {\bf 1101} (2011) 016
  [arXiv:1008.5157 [hep-ph]].
  

\bibitem{DeSimone:2012qr}
  A.~De Simone and A.~Riotto,
  arXiv:1208.1344 [hep-ph].


\bibitem{Dvali:2003em}
  G.~Dvali, A.~Gruzinov and M.~Zaldarriaga,
  Phys.\ Rev.\  D {\bf 69} 023505 (2004).

  
\bibitem{Lyth:2005qk}
  D.~H.~Lyth,
  JCAP {\bf 0511 } (2005)  006.
  [astro-ph/0510443].

\bibitem{Lyth:2006nx}
  D.~H.~Lyth and A.~Riotto,
  Phys.\ Rev.\ Lett.\  {\bf 97} (2006) 121301
  [astro-ph/0607326].


\bibitem{Komatsu:2010fb}
  E.~Komatsu {\it et al.},
   Astrophys.\ J.\ Suppl.\  {\bf 192}, 18 (2011).


\bibitem{EliasMiro:2011aa}
  J.~Elias-Miro, J.~R.~Espinosa, G.~F.~Giudice, G.~Isidori, A.~Riotto and A.~Strumia,
  Phys.\ Lett.\ B {\bf 709} (2012) 222
  [arXiv:1112.3022 [hep-ph]].


\bibitem{Bezrukov:2012sa}
  F.~Bezrukov, M.~Y.~.Kalmykov, B.~A.~Kniehl and M.~Shaposhnikov,
  JHEP {\bf 1210} (2012) 140
  [arXiv:1205.2893 [hep-ph]].

\bibitem{Degrassi:2012ry}
  G.~Degrassi, S.~Di Vita, J.~Elias-Miro, J.~R.~Espinosa, G.~F.~Giudice, G.~Isidori and A.~Strumia,
  JHEP {\bf 1208} (2012) 098
  [arXiv:1205.6497 [hep-ph]].


\bibitem{Mollerach:1989hu}
  S.~Mollerach,
  Phys.\ Rev.\  D {\bf 42} 313 (1990).

\bibitem{Linde:1996gt}
  A.~D.~Linde and V.~F.~Mukhanov,
  Phys.\ Rev.\  D {\bf 56} 535 (1997).

\bibitem{CurvatonLW}
  D.~H.~Lyth and D.~Wands,
  Phys.\ Lett.\  B {\bf 524}, 5 (2002).


\bibitem{CurvatonMT}
  T.~Moroi and T.~Takahashi,
  Phys.\ Lett.\  B {\bf 522}, 215 (2001)
  [Erratum-ibid.\  B {\bf 539}, 303 (2002)].

\bibitem{CurvatonES}
  K.~Enqvist and M.~S.~Sloth,
  Nucl.\ Phys.\  B {\bf 626}, 395 (2002).


\bibitem{CurvatonSelf}
  K.~Dimopoulos, G.~Lazarides, D.~Lyth and R.~Ruiz de Austri,
  Phys.\ Rev.\  D {\bf 68} 123515 (2003); \\
  K.~Enqvist and S.~Nurmi,
  JCAP {\bf 0510} 013 (2005); \\
  Q.~G.~Huang,
  Phys.\ Lett.\ B {\bf 669}, 260 (2008);\\
  K.~Enqvist and T.~Takahashi,
  JCAP {\bf 0809}, 012 (2008); \\
  Q.~G.~Huang and Y.~Wang,
  JCAP {\bf 0809} 025 (2008); \\
  Q.~G.~Huang,
  JCAP {\bf 0811}, 005 (2008); \\
  M.~Kawasaki, K.~Nakayama and F.~Takahashi,
  JCAP {\bf 0901}, 026 (2009); \\
  P.~Chingangbam and Q.~G.~Huang,
  JCAP {\bf 0904}, 031 (2009); \\
  K.~Enqvist, S.~Nurmi, G.~Rigopoulos, O.~Taanila and T.~Takahashi,
  JCAP {\bf 0911}, 003 (2009); \\
  K.~Enqvist and T.~Takahashi,
  JCAP {\bf 0912}, 001 (2009); \\
  K.~Enqvist, S.~Nurmi, O.~Taanila and T.~Takahashi,
  JCAP {\bf 1004}, 009 (2010); \\
  Q.~G.~Huang,
  JCAP {\bf 1011}, 026 (2010)
  [Erratum-ibid.\  {\bf 1102}, E01 (2011)]; \\
  C.~T.~Byrnes, K.~Enqvist and T.~Takahashi,
  JCAP {\bf 1009}, 026 (2010); \\
  K.~-Y.~Choi and O.~Seto,
  Phys.\ Rev.\ D {\bf 82} 103519 (2010); \\
  J.~Fonseca and D.~Wands,
  Phys.\ Rev.\  D {\bf 83}, 064025 (2011); \\
  C.~T.~Byrnes, K.~Enqvist, S.~Nurmi and T.~Takahashi,
  JCAP {\bf 1111}, 011 (2011); \\
  M.~Kawasaki, T.~Kobayashi and F.~Takahashi,
  Phys.\ Rev.\  D {\bf 84}, 123506 (2011)
  [Phys.\ Rev.\  D {\bf 85}, 029905 (2012)];\\
  A.~Mazumdar and J.~Rocher,
  Phys.\ Rept.\  {\bf 497}, 85 (2011)
  [arXiv:1001.0993 [hep-ph]].

\bibitem{MixedIC}
  D.~Langlois and F.~Vernizzi,
  Phys.\ Rev.\  D {\bf 70}, 063522 (2004); \\
  G.~Lazarides, R.~R.~de Austri and R.~Trotta,
  Phys.\ Rev.\  D {\bf 70}, 123527 (2004); \\
  F.~Ferrer, S.~Rasanen and J.~Valiviita,
  JCAP {\bf 0410}, 010 (2004); \\
  T.~Moroi, T.~Takahashi and Y.~Toyoda,
  Phys.\ Rev.\  D {\bf 72} 023502 (2005);\\
  T.~Moroi and T.~Takahashi,
  Phys.\ Rev.\  D {\bf 72} 023505 (2005); \\
  K.~Ichikawa, T.~Suyama, T.~Takahashi and M.~Yamaguchi,
  Phys.\ Rev.\  D {\bf 78}, 023513 (2008);\\
  K.~-Y.~Choi and O.~Seto,
  Phys.\ Rev.\ D {\bf 85} (2012) 123528.


\bibitem{CurvatonNG}
  D.~H.~Lyth, C.~Ungarelli and D.~Wands,
  Phys.\ Rev.\  D {\bf 67}, 023503 (2003),  
  N.~Bartolo, S.~Matarrese and A.~Riotto,
  Phys.\ Rev.\  D {\bf 69} 043503 (2004),
  D.~H.~Lyth and Y.~Rodriguez,
  Phys.\ Rev.\ Lett.\  {\bf 95} 121302 (2005),
  M.~Sasaki, J.~Valiviita and D.~Wands,
  Phys.\ Rev.\  D {\bf 74} 103003 (2006),
 K.~A.~Malik and D.~H.~Lyth,
  JCAP {\bf 0609} 008 (2006),
  K.~Y.~Choi and J.~O.~Gong,
  JCAP {\bf 0706} 007 (2007).








\bibitem{Bezrukov:2008ut}
  F.~Bezrukov, D.~Gorbunov and M.~Shaposhnikov,
  JCAP {\bf 0906} (2009) 029
  [arXiv:0812.3622 [hep-ph]].
\bibitem{GarciaBellido:2008ab}
  J.~Garcia-Bellido, D.~G.~Figueroa and J.~Rubio,
  Phys.\ Rev.\ D {\bf 79} (2009) 063531
  [arXiv:0812.4624 [hep-ph]].




\bibitem{Kunimitsu:2012xx}
  T.~Kunimitsu and J.~'i.~Yokoyama,
  arXiv:1208.2316 [hep-ph].



  
\bibitem{Kofman:2003nx}
  L.~Kofman,
  arXiv:astro-ph/0303614.

\bibitem{Zaldarriaga:2003my}
  M.~Zaldarriaga,
  Phys.\ Rev.\  D {\bf 69}, 043508 (2004).

\bibitem{Alabidi:2010ba}
  L.~Alabidi, K.~Malik, C.~T.~Byrnes and K.~-Y.~Choi,
  JCAP {\bf 1011} (2010) 037
  [arXiv:1002.1700 [astro-ph.CO]].




\bibitem{Ichikawa:2008ne}
  K.~Ichikawa, T.~Suyama, T.~Takahashi, M.~Yamaguchi,
  Phys.\ Rev.\  {\bf D78 }  063545 (2008).

    

\bibitem{Planck} 
  http://www.rssd.esa.int/index.php?project=PLANCK


\end{thebibliography}
\end{document}